\documentclass[11pt,twoside]{article}
\usepackage{asp2010}

\usepackage{graphicx}

\resetcounters

\begin{document}

\title{Astrometric determination of  WD  radial velocities with Gaia?}
\author{Stefan Jordan$^1$, 
    Jos de Bruijne$^2$}
\affil{$^1$Astronomisches Rechen-Institut, Zentrum f\"ur Astronomie der Universit\"at Heidelberg, M\"onchhofstr. 12--14, D-69120 Heidelberg, Germany}
\affil{$^2$European Space Research and Technology Centre (ESA/ESTEC) in Noordwijk, the Netherlands}

\section{Introduction}

Usually, the determination of radial velocities of stars relies on the shift of spectral lines by the Doppler effect. \citet{RusselAtkinson:1931} and
\citet{Oort:1932} already noted that due to the large proper motion and parallax of the white dwarf (WD) van Maanen 2, a determination of the perspective
acceleration of the proper motion would provide a direct astrometric determination of the radial velocity which is independent of the gravitational redshift.
If spectroscopic redshift measurements of H$_\alpha$ and H$_\beta$ NLTE cores  exist, a purely
astrometric determination would allow disentangling the gravitational redshift from the Doppler shift.

The best instrument for measuring the tiny perspective acceleration is the Gaia satellite of the European Space Agency, aiming at absolute astrometric
measurements of one billion stars down to $20^{\rm th}$ magnitude with unprecedented accuracy. At $15^{\rm th}$ magnitude, the predicted angular accuracy of
Gaia is $\sim$20 micro-arcseconds ($\mu$as). In this article, we estimate whether it is possible to measure the radial velocity of WDs
astrometrically by the exchange of proper motion into radial velocity during the 5-year mission of the satellite or by combining Hipparcos data with the
position measurements at the beginning of the Gaia mission (the Hundred-Thousand-Proper-Motion project HTPM). 

\section{Perspective acceleration}
\begin{center}
\begin{figure}
\centering
\includegraphics[type=eps,ext=.eps,read=.eps,width=0.35\textwidth]{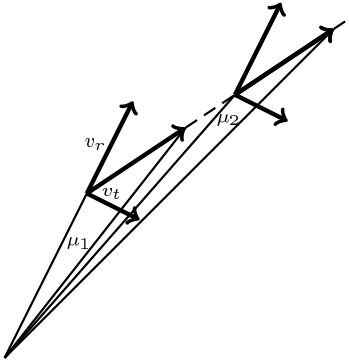}
\caption{Perspective acceleration: let $\mu_1$ be the proper motion at the beginning of the Gaia mission and $\mu_2$ the proper motion at the end of the mission. As a (nearby) star passes the observer (Sun), proper motion is continuously exchanged into radial velocity. Measuring the decrease of the proper motion allows astrometric determination of the radial velocity, without the use of spectroscopy.}
\label{Fig:perspective}
\end{figure}
\end{center}

Since stars move with (almost) constant velocity through space, the proper motion $\mu=(\mu_{\alpha\ast}^2+\mu_\delta^2)^{1/2}$ as seen from an observer varies
inversely with the distance to the object and also changes due to the varying angle between the line of sight and the space-velocity vector \cite[see,
e.g.,][see also Fig.\,\ref{Fig:perspective}]{Dravins-etal:1999}. Both effects lead to an apparent change in proper motion of $\dot{\mu} = \frac{-2 \mu \pi
v_r}{1\,{\rm AU}}$ \citep{vandeKamp:1967,Murray:1983}, where $\pi$ denotes the parallax of the star. Conversely, this means that the radial velocity can be
determined astrometrically from $
v_r = -\frac{\dot{\mu} {\rm \ 1\,AU}}{2\pi \mu}.
$

As summarized by \cite{Dravins-etal:1999}, past attempts to measure perspective acceleration yielded barely significant or spurious results. A high accuracy in terms of the astrometric radial velocity was only obtained for Barnard's star \citep{vandeKamp:1981} and for van Maanen 2 \citep{GatewoodRussel:1974}. By combining positions and proper motions from Hipparcos and old position measurements from the Astrographic Catalogue \cite[for a description, see][]{Eichhorn:1974}, \citet{Dravins-etal:1999} determined astrometric radial velocities for sixteen stars, albeit with modest precisions.

\section{The Hundred-Thousand-Proper-Motion project}

The HTPM project is part of the first intermediate release of the Gaia data and will contain the proper motions of $\sim
113,500$ stars using an $\sim 23$-year baseline. The proper motions will be based on the Hipparcos data (positions, proper motions, and parallaxes) at the first
epoch (1991.25) and Gaia positional data at the second epoch ($\sim 2014.25$). For this catalogue, \cite{deBruijneEilers12} determined that for nearly 100
stars, the radial velocities available in the literature are insufficiently precise to correct for the perspective acceleration. Generally, stars with a large
product $\mu \pi v_r$, i.e., nearby, fast-moving stars, are good candidates for measuring radial velocities through astrometry.

\begin{table}[h]
\caption{Predicted standard errors of astrometric radial velocities based on HTPM data and on Gaia data for 39 WDs (all Hipparcos entries of white
dwarfs themselves plus all Hipparcos stars with known white-dwarf companions; HIP35307 has been omitted since it has a negative parallax measurement). Columns:
1: Hipparcos identifier;
2: $Hp$ magnitude;
3: parallax in mas from the XHIP Catalogue \citep{AndersonFrancis12};
4: parallax error in mas (XHIP);
5: proper motion in $\mu$as yr$^{-1}$ (XHIP);
6: literature radial velocity in km s$^{-1}$ (XHIP);
7: literature radial-velocity error in km s$^{-1}$ (XHIP); footnote: quality flag (A = good, ..., D = bad; XHIP);
8: perspective acceleration in $\mu$as yr$^{-2}$;
9: accumulated effect of perspective acceleration on position over 23 yr in $\mu$as (the HTPM temporal baseline);
10: HTPM astrometric radial velocity standard error in km s$^{-1}$;
11: accumulated effect of perspective acceleration on position over 2.5 yr in $\mu$as (half the Gaia mission timeline);
12: Gaia astrometric radial velocity standard error in km s$^{-1}$.}
\begin{center}
{\scriptsize
\begin{tabular}{
p{0.6cm}p{0.4cm}p{0.7cm}p{0.5cm}p{0.7cm}rp{0.4cm}rrrrrrrrrrrrrrr
}
 \multicolumn{1}{c}{1}     &  \multicolumn{1}{c}{ 2}  &      \multicolumn{1}{c}{ 3}   &     \multicolumn{1}{c}{ 4 } &    \multicolumn{1}{c}{ 5 } &   
\multicolumn{1}{c}{ 6} &   \multicolumn{1}{c}{  7}  & \multicolumn{1}{c}{ 8} &   \multicolumn{1}{c}{ 9 }
&       \multicolumn{1}{c}{ 10}    &    \multicolumn{1}{c}{   11}   &    \multicolumn{1}{c}{    12  } \\
\hline \\
002600&10.42&11.22&1.52&227.21&49.98&0.17$^{\rm A}$&
-0.26&-68.93&$3\cdot 10^5$&-0.81&
1040\\
003829&12.56&234.60&5.90&2978.19&263.00&4.90$^{\rm D}$&
-375.85&$-10^5$&444&-1175&
3.8\\
008709&12.55&64.53&3.40&122.18&64.00& &
-1.03&-273&23738&-3.23&
336\\
011650&12.85&37.52&5.17&75.98&-5.70&2.90$^{\rm D}$&
0.03&8.79&191848&0.10&
930\\
012031&12.37&10.90&3.94&83.22&55.20&7.40$^{\rm D}$&
-0.10&-27.09&149151&-0.32&
2922\\
014754&11.44&98.50&1.85&109.00&33.80&3.20$^{\rm D}$&
-0.74&-196.33&12007&-2.32&
247\\
018824&6.86&19.35&0.63&373.70&10.21&0.20$^{\rm A}$&
-0-15&-39.94&24697&-0.47&
367\\
021088&10.79&179.27&3.23&2426.09&27.90&0.40$^{\rm A}$&
-24.82&-6564.89&570&-77.56&
6.1\\
021482&8.23&56.02&1.43&276.52&36.02&0.08$^{\rm A}$&
-1.14&-301.87&4415&-3.57&
171\\
023692&11.72&16.70&2.97&91.90&69.00& &
-0.22&-57.29&94943&-0.68&
1727\\
027878&7.88&18.68&0.81&260.59&16.10&15.60$^{\rm B}$&
-0.16&-42.40&19235&-0.50&
545\\
029788&6.56&26.72&0.29&319.81&-1.50&0.67$^{\rm A}$&
0.03&6.93&3575&0.08&
310\\
032560&12.09&63.53&3.55&967.42&80.00&5.00$^{\rm D}$&
-10.06&-2660.08&2860&-31.43&
43\\
037853&5.48&65.75&0.51&1709.12&106.16&0.10$^{\rm A}$&
-24.40&-6454.15&833&-76.25&
24\\
054530&8.82&24.90&0.98&234.45& & & 
0.03&6.98&24466&0.08&
454\\
056662&12.53&63.26&3.60&147.94&17.00&42.00$^{\rm D}$&
-0.33&-86.07&18193&-1.02&
283\\
057367&11.59&217.01&2.40&2687.68& & & 
2.64&697.36&282&8.24&
4.5\\
059519&10.10&22.18&1.49&428.69& & & 
0.04&11.37&15708&0.13&
279\\
064766&12.67&25.96&6.38&192.62&44.40&14.40$^{\rm A}$&
-0.45&-120.11&96111&-1.42&
530\\
065877&12.39&55.50&3.85&1205.95&35.60&4.70$^{\rm A}$&
-4.87&-1289.08&2678&-15.23&
40\\
066578&12.84&38.29&3.02&404.25&1.00& &
-0.03&-8.37&17732&-0.10&
171\\
073224&9.98&16.51&1.66&208.09& & & 
0.02&4.11&37400&0.05&
772\\
077358&6.15&65.13&0.40&467.30&-6.70&0.74$^{\rm A}$&
0.42&110.32&1008&1.30&
87\\
080300&10.99&76.00&2.56&76.45&13.00&0.10$^{\rm D}$&
-0.15&-40.87&22075&-0.48&
456\\
080522&10.20&17.86&2.12&492.02&90.70&2.70$^{\rm A}$&
-1.63&-431.20&15537&-5.09&
302\\
082257&12.38&94.04&2.67&326.13&41.60& &
-2.61&-690.24&6596&-8.16&
86\\
092306&9.80&12.68&0.76&182.50& & & 
0.01&2.77&27645&0.03&
1146\\
095071&12.37&91.31&4.02&173.58& & & 
0.07&18.95&10385&0.22&
167\\
101516&11.55&64.32&2.58&693.46&71.00&7.40$^{\rm D}$&
-6.48&-1713.31&2495&-20.24&
60\\
102207&12.42&48.22&3.77&367.77&1.80& &
-0.07&-17.27&12129&-0.20&
149\\
102488&2.64&44.86&0.12&484.93&-11.60&0.30$^{\rm A}$&
0.52&136.52&1754&1.61&
122\\
103393&11.91&56.54&3.92&818.64&-42.70&1.00$^{\rm A}$&
4.04&1069.26&8565&12.63&
57\\
106335&9.86&20.26&2.01&413.78&-109.60&0.40$^{\rm A}$&
1.88&497.09&11240&5.87&
316\\
107968&12.88&37.51&4.41&302.18&32.00& &
-0.74&-196.23&13009&-2.32&
234\\
110218&10.30&20.30&1.40&420.92& & & 
0.04&10.22&13211&0.12&
310\\
113231&8.14&27.22&1.12&552.73&-27.68&0.14$^{\rm A}$&
0.85&225.31&6233&2.66&
176\\
113244&11.30&40.89&2.12&447.31&-1.00&10.00$^{\rm D}$&
0.04&9.90&7496&0.12&
145\\
113786&8.78&14.97&0.79&160.19&13.63& &
-0.07&-17.68&35508&-0.21&
1105\\
117308&11.40&19.11&3.04&300.40& & & 
0.03&6.86&32444&0.08&
462\\
\end{tabular}}
\end{center}
\label{tab:summary}
 
\end{table}

\section{Perspectives for white dwarfs}

We estimated the accuracy of astrometric $v_r$ determinations with Gaia by investigating the 39 WDs either having
Hipparcos astrometry themselves \citep{Vauclair-etal:1997} or wide companions with Hipparcos measurements \citep{GouldChaname:2004}.

First, we estimate the accuracy using HTPM data only. In this case, we have positions $\phi_{1991.25}$ and proper motions $\mu_{1991.25}$ determined at epoch
1991.25 and only positions $\phi_{2014.25}$ from Gaia at $\sim 2014.25$. $\phi_{2014.25}$ combined with the $\phi_{1991.25}$ positions provide a
second proper-motion measurement (symbolically: $\mu = (\phi_{2014.25} - \phi_{1991.25}) / T$). The standard error of  the astrometric $v_r$ can
be estimated by \citep{Dravins-etal:1999}
\begin{equation}
\epsilon(v_r) \approx \frac{2 \cdot {\rm \ 1 AU}}{T \pi \mu_{1991.25}}\left[\frac{\epsilon(\phi_{1991.25})^2+\epsilon(\phi_{2014.25})^2}{T}+\epsilon(\mu_{1991.25})^2\right]^{1/2},
\end{equation}
with $T = 2014.25 - 1991.25$ and $\epsilon(\mu)$ the accuracy of the proper-motion measurements. The expected HTPM proper-motion standard errors are
$\sim$30--190 $\mu$as~yr$^{-1}$, depending on magnitude. For our 39-star sample, the best accuracy for the astrometric radial velocity using only HTPM data is
282 km~s$^{-1}$ for HIP57367 (column 10 in Table\,\ref{tab:summary}), which is too large to be of any scientific value.

Second, we estimate the accuracy using the Gaia data only. The achievable accuracy for quasi-continuous observations during a period of length $L$ (5 years in
the case of Gaia) is given by \citep{Dravins-etal:1999}:
\begin{equation}
\epsilon(v_r) \approx \sqrt{15} \cdot \frac{\epsilon({\mu}) {\rm \ 1 AU}}{L\pi \mu}.
\end{equation}

Since all stars in our sample are bright for Gaia, the end-of-mission accuracy $\epsilon(\mu)$ amounts to $\sim$$3.5\,\mu$as~yr$^{-1}$
\citep{deBruijne:2012b}. For this case, three stars (HIP 3829 = van Maanen 2, 21088 = Gl 169.1B, and 57367 = GJ 440) have predicted standard errors below 10
km~s$^{-1}$ (column 12 in Table\,\ref{tab:summary}). Five objects have predicted standard errors between 20 and 60 km~s$^{-1}$, which is of the order of the
mean gravitational redshift in DA WDs, which is about 33 km~s$^{-1}$ \citep{Falcon-etal:2010}.

\section{Conclusions}
An astrometric determination of radial velocities of WDs with Gaia data is possible but limited to a few bright, nearby objects.
The accuracy of the HTPM catalogue, which will be released early during the Gaia mission, is by far not large enough for an astrometric radial velocity
determination of WDs. A breakthrough in this area would require a second micro-arcsecond astrometry mission, the data of which would have to be
combined with the Gaia data \citep[see, e.g., ][]{2010IAUS..261..342A}.

\bibliography{Jordan_deBruijne_arXiv.bbl}

\end{document}